\documentclass[preprint,prd,nofootinbib,tightenlines,amsmath]{revtex4}

\usepackage{epsfig}
\usepackage{graphics,color}      
\usepackage{verbatim}
\usepackage{amsmath}    
\catcode`\@=11
\def\sla@#1#2#3#4#5{{%
 \setbox\z@\hbox{$\m@th#4#5$}%
 \setbox\tw@\hbox{$\m@th#4#1$}%
 \dimen4\wd\ifdim\wd\z@<\wd\tw@\tw@\else\z@\fi
 \dimen@\ht\tw@
 \advance\dimen@-\dp\tw@ \advance\dimen@-\ht\z@
 \advance\dimen@\dp\z@
 \divide\dimen@\tw@ \advance\dimen@-#3\ht\tw@
 \advance\dimen@-#3\dp\tw@ \dimen@ii#2\wd\z@
 \raise-\dimen@\hbox to\dimen4{%
 \hss\kern\dimen@ii\box\tw@\kern-\dimen@ii\hss}%
 \llap{\hbox to\dimen4{\hss\box\z@\hss}}}}

\def\slashed#1{%
 \expandafter\ifx\csname sla@\string#1\endcsname\relax
{\mathpalette{\sla@/00}{#1}}
\fi}
\def\declareslashed#1#2#3#4#5{%
 \expandafter\def\csname sla@\string#5\endcsname{%
#1{\mathpalette{\sla@{#2}{#3}{#4}}{#5}}}}
 \catcode`\@=12
\declareslashed{}{/}{.08}{0}{D}
 \declareslashed{}{/}{.1}{0}{A}
 \declareslashed{}{/}{0}{-.05}{k}
 \declareslashed{}{/}{.1}{0}{\partial}
 \declareslashed{}{\not}{-.6}{0}{f}

\def\lsim{\mathrel {\vcenter {\baselineskip 0pt \kern 0pt
    \hbox{$<$} \kern 0pt \hbox{$\sim$} }}}
\def\gsim{\mathrel {\vcenter {\baselineskip 0pt \kern 0pt
    \hbox{$>$} \kern 0pt \hbox{$\sim$} }}}
\def\h{{\lambda_{12}}}

\begin{document}

\baselineskip=15pt

\preprint{hep-ph/0611085}

\hspace*{\fill} $\hphantom{-}$

\title{The role of $D^{\star\star}$ in $B^-\to D_s^+ K^- \pi^-$}

\author{Oleg Antipin and G. Valencia}

\email{oaanti02@iastate.edu}
\email[]{valencia@iastate.edu}

\affiliation{Department of Physics and Astronomy, Iowa State University, Ames, IA 50011\\}

\date{\today}

\begin{abstract}

The BaBar collaboration has recently reported the observation of the decay mode $B^-\to D_s^+ K^- \pi^-$. We investigate the role played by the $D^{\star\star}$ resonances in this decay mode using HQET. Although these resonances cannot appear as physical intermediate states in this reaction, their mass is very close to the $D_s^+ K^-$ production threshold and may, therefore, play a prominent role. We pursue this possibility to extract information on the properties of the strong $D^{\star\star} D M$ couplings. As a byproduct of this analysis we point out that future super-$B$ factories may be able to measure the $D_0^0 D^\star \gamma$ radiative coupling through the reaction $B^-\to D^\star \gamma \pi^-$. 

\end{abstract}

\pacs{PACS numbers: }

\maketitle

\section{Introduction}

The BaBar collaboration has recently reported the observation of the decay mode $B^-\to D_s^+ K^- \pi^-$  with a branching ratio ${\cal B}(B^-\to D_s^+ K^- \pi^-) =(1.88\pm 0.13\pm0.41) \times  10^{-4} $~\cite{batalk}. This decay mode is different from  the mode $B^-\to D^{\star\star}\pi^-\to D^+ \pi^- \pi^-$ observed by Belle \cite{Abe:2003zm} in that the $D^{\star\star}$ resonances are too light to decay into $D_s^+ K^-$. Nevertheless their masses \cite{Abe:2003zm}, 
\begin{eqnarray}
m_{D_0^{\star\star}}&=&(2308 \pm 17\pm15\pm 28)~MeV  \nonumber \\ 
m_{D_2^{\star\star}}& =&(2461.6 \pm 2.1\pm 0.5 \pm 3.3)~MeV, 
\label{measmass}
\end{eqnarray}
are sufficiently close to the threshold for production of $D_s^+ K^-$ that we can entertain the possibility of them playing a significant role in $B^-\to D_s^+ K^- \pi^-$ as  ``quasi-resonant'' intermediate states. 

In this paper we use heavy quark effective theory (HQET) to investigate this possibility. This study will serve as a probe of the properties of the $D^{\star\star} D M$ interactions, where $M$ a member of the light pseudoscalar meson octet. In particular we can check the $SU(3)$ relations in strong $D^{\star\star}$ decay. In addition, an analysis of a distribution with respect to the angle between the pion and kaon momenta can further constrain the $D_2^0$ tensor couplings. 

\begin{figure}[htb]
\includegraphics[width=6in]{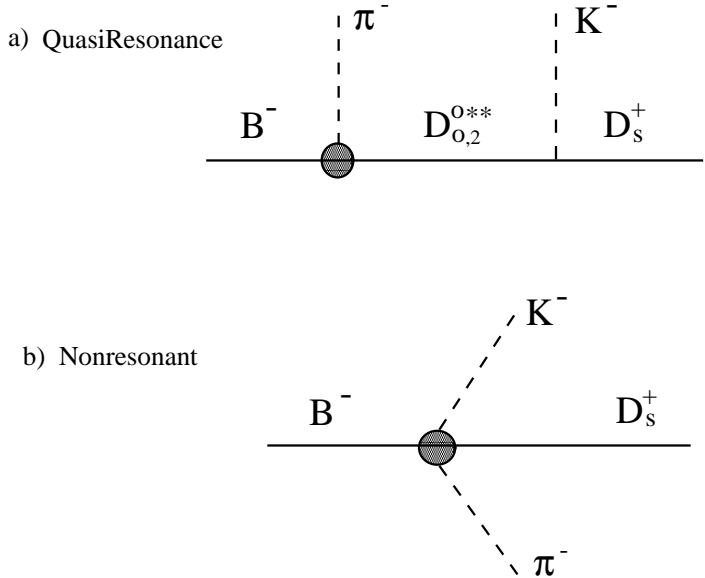}
\centering
\caption{Decomposition of the decay mode $B^-\to D_s^+ K^- \pi^-$ into contributions that are mediated by a $D^{\star\star}$ that is near its mass shell and those that are not.}
\label{blob}
\end{figure}

Schematically, our procedure consists of  splitting the decay $B^-\to D_s^+ K^- \pi^-$ into ``quasi-resonant'' and non-resonant contributions as depicted in Figure~\ref{blob}. 
If the $D^{\star\star}$ resonances were heavy enough to decay into $D_s^+ K^-$ we would expect the ``quasi-resonant'' contribution to dominate. Furthermore, in the narrow width approximation the production and decay processes would factorize, and we could study the properties of the strong decay vertex. We investigate the extent to which the ``quasi-resonant'' process dominates by first computing the amplitudes with the aid of heavy quark effective theory (HQET). We then normalize the resulting rates to the two-body $B^- \to D_{0,2}^0 \pi^-$ weak decay rates and use this as a constraint on the weak transition. Finally we study the behavior of the normalized rates for different parametrizations of the weak vertex, treating the residual dependence on the weak vertex as an indication of the extent to which the ``quasi-resonant'' contribution dominates. 
 
\section{Formalism}

We use the HQET formalism to describe the interactions involving the heavy meson $(0^-,1^-)$ doublet, its excited positive parity partners $(0^+,1^+)$ and $(1^+,2^+)$, and light 
pseudo-scalar mesons \cite{Wise:1992hn,Yan:1992gz,Burdman:1992gh,Falk:1991nq,manoharbook,Casalbuoni:1996pg}. We follow standard notation to incorporate the light pseudo-scalars as the Goldstone bosons of spontaneously broken chiral symmetry through the matrix 
$\xi=\exp (\frac{iM}{f_{\pi}})$ with a normalization in which the pion decay constant is $f_{\pi}=132$~MeV. The matrix $M$ is explicitly given by,
\begin{equation}
M= \begin{pmatrix} \sqrt{\frac{1}{2}}\pi^0+\sqrt{\frac{1}{6}}\eta & \pi^+ & K^+ \\ \pi^- & -\sqrt{\frac{1}{2}}\pi^0+\sqrt{\frac{1}{6}}\eta  & K^0 \\  K^- & \bar{K}^0 & -\sqrt{\frac{2}{3}}\eta  \end{pmatrix} .
\label{pionmatrix}
\end{equation}
Similarly, the heavy meson doublets are described by the following fields and their conjugates,
\begin{eqnarray}
(0^-,1^-) &\to & H=\frac{1+\slashed{v}}{2}(\slashed{P^*}-P\gamma_5),\,  \bar{H}=\gamma_0 H^{\dagger} \gamma_0\nonumber \\
(0^+,1^+) &\to & S=\frac{1+\slashed{v}}{2}(\slashed{P_1}\gamma_5-P_0), \, \bar{S}=\gamma_0 S^{\dagger} \gamma_0 \nonumber \\
(1^+,2^+)&\to & T^{\mu}=\frac{1}{2}(1+\slashed{v})\left[P_2^{\mu\nu}\gamma_{\nu}-\sqrt{3/2}\tilde{P}_{1\nu}\gamma_{5}(g^{\mu\nu}-\frac{1}{3}
\gamma^{\nu}(\gamma^{\mu}-v^{\mu})) \right], \nonumber \\
&& \bar T^{\mu}=\gamma_0 T^{\dagger\mu}\gamma_0.
\end{eqnarray}

At leading order in the heavy quark and chiral expansions, the strong interaction mediated decays of the form ${H,S,T}\to H M$ are described by the Lagrangians \cite{Wise:1992hn,Yan:1992gz,Burdman:1992gh,Kilian:1992hq}
\begin{eqnarray}
{\cal L}_{H}& =& g  \, \, Tr\left[ H \gamma_{\mu}\gamma_{5}A^{\mu}\bar{H}\right], \nonumber \\
{\cal L}_S &=& h \,\, Tr\left[ S\gamma_{\mu}\gamma_{5}A^{\mu}\bar{H}\right]+{\rm h.c.}\, , \nonumber \\
{\cal L}_T &=& \frac{h_1}{\Lambda_{\chi}}  Tr\left[H\gamma_{\lambda}\gamma_{5}(D_{\mu}A^{\lambda})\bar{T}^{\mu}\right]+ \frac{h_2}{\Lambda_{\chi}} Tr\left[H \gamma_{\lambda}\gamma_{5}(D^{\lambda}A_{\mu})\bar{T}^{\mu}\right]+{\rm h.c.},
\label{strongl}
\end{eqnarray}
where the axial current is given by,
\begin{equation}
A_{\mu}\equiv \frac{i}{2}(\xi^{\dagger}\partial_{\mu}\xi-\xi \partial_{\mu} \xi^{\dagger}),
\end{equation}
and the traces are over Dirac and flavor indices. For our numerical estimates we will use the values  $|h^\prime|=|(h_1+h_2)/\Lambda_{\chi}|\approx 0.5$~GeV$^{-1}$, $h=-0.52$ and $g=0.4$ 
\cite{Falk:1992cx,Casalbuoni:1996pg}. We will also replace $f_\pi \to f_K \sim 1.3 f_\pi$  where appropriate.

It is a simple exercise to write the corresponding weak vertices describing the transitions from a $b$-quark meson to a $c$-quark meson. In this case, however, we do not expect a reliable description of the weak transition as the $m_b-m_c$ mass difference is larger than $\Lambda_{\chi}$. We will use the HQET framework to parametrize the weak transitions  in a manner similar to that of Ref.~\cite{Jugeau:2005yr}. We then treat the result as a phenomenological description of the weak transition in terms of three free parameters that are constrained by the two body decays $B^- \to D_{0,2}^0 \pi^-$. 

The dominant short distance operator responsible for the decays $B^-\to D_s^+ K^- \pi^-$, $B^-\to D^+ \pi^- \pi^-$ is an $SU(3)$ octet of the form 
$\bar{c}\gamma_{\mu}(1-\gamma_5)b\bar{d}\gamma^{\mu}(1-\gamma_5)u$. We use standard techniques \cite{Georgi:1985kw} to introduce this operator into the HQET formalism. We first construct the matrix $\h$ with  
$\h^i_j=\delta^i_1 \delta^2_j$ to represent the $SU(3)$ properties of the operator. We then pretend that $\h$ transforms as $\h \to L\h L^{\dagger}$ under chiral symmetry and construct chiral symmetric operators that include $\h$. The transformation properties under chiral symmetry of the other relevant objects are $H_Q \to H_Q U^{\dagger}$,$\hspace{1mm}\bar{H}_Q \to U\bar{H}_Q$, $\hspace{1mm}\xi \to L\xi U^{\dagger}$ and $\hspace{1mm}\xi^{\dagger} \to U \xi^{\dagger} L^{\dagger}$. With these ingredients we construct the effective weak Lagrangian beginning with the $H_b \to H_c$ transitions. There is only one term without derivatives (the sign is chosen to match the notation in \cite{Antipin:2006dn}),
\begin{eqnarray}
{\cal L}_W &=& 
\beta_W^\prime Tr\left[\hspace{1mm}H_{b}\xi^{\dagger } \hspace{1mm}\gamma_{\mu}(1-\gamma_5) \hspace{1mm}\h \hspace{1mm}\xi \bar{H}^{\bar{c}}\hspace{1mm} \gamma^{\mu}(1-\gamma_5)\hspace{1mm}\right].\nonumber\\
\label{neutral} 
\end{eqnarray}
There is also a unique term with one derivative,
\begin{equation}
{\cal L}_{W1} = ik_H \hspace{1mm} Tr\left[H_{bj}  \bar{H}^{\bar{c}j} \gamma_{\mu}(1-\gamma_{5}) \right] \hspace{2mm} Tr\left[\xi^{\dagger}\h\partial^{\mu}\xi \right].
\label{der} 
\end{equation}
Even though the operator Eq.~\ref{der} is formally suppressed by one order in the momentum expansion with respect to the one in Eq.~\ref{neutral} we will keep both of them for several reasons. First, because in a matching to the  underlying quark-operator Eq.~\ref{neutral} is suppressed by $1/N_c$ with respect to Eq.~\ref{der}; second, because we do not really expect the momentum expansion to be relevant in $b\to c$ transitions; and third, because Ref.~\cite{Jugeau:2005yr} finds that an interplay of these two operators is necessary to describe the $B^-\to D^+ \pi^- \pi^-$ decay. For all these reasons we will limit our discussion of the weak transitions to these two operators. In particular we will also ignore small contributions proportional to $V_{ub}$, as well as perturbative QCD corrections to the Wilson coefficient of the quark operator. 

For weak transitions of the form $H_b \to S_c$ we consider two weak operators analogous to Eqs.~\ref{neutral}~and~\ref{der} obtained by replacing $H_c\to S_c$. In principle these operators have different coefficients than those in  Eqs.~\ref{neutral}~and~\ref{der}. However, we  estimate all the weak coefficients  in naive factorization where  the constant $\beta_W^\prime$  for $H_b\to S_c$ transitions is related to the constant  $\beta_W^\prime$ for $H_b\to H_c$ transitions by the factor $(f_{D_0}\sqrt{m_{D_0}})/(f_D\sqrt{m_D})$. For our estimates we take this factor to be one. Several models to estimate the quantities in this ratio are discussed in Ref.~\cite{Jugeau:2005yr}, resulting in the ratio differing from one by factors of at most two. In all the models considered in Ref.~\cite{Jugeau:2005yr} the two constants have the same sign. 

For weak transitions involving a $T$ field, Eq.~\ref{neutral} will not have an analogue in this approximation (the tensor decay constant vanishes \cite{Casalbuoni:1996pg,Veseli:1996yg,Neubert:1997hk,LeYaouanc:1996bd}). The single weak operator in this case reads,
\begin{equation}
{\cal L}_T=i k_T Tr\left [v_{\alpha}\bar{T}^{\alpha}_{v^\prime}\gamma_{\mu}(1-\gamma_5)H_v] \hspace{2mm}Tr[\xi^{\dagger}\h\partial^{\mu}\xi\right].
\label{tenoper} 
\end{equation}
The $k_H$, $k_S$ and $k_T$ coefficients are related in factorization, with their relative signs being given by those of 
the Isgur-Wise functions \cite{Isgur:1990jf} $\xi(\omega), \tau_{1/2,3/2}(\omega)$. The factorization results that we use are,
\begin{eqnarray}
\beta_W^{\prime}&=&\frac{G_F V_{cb} V_{ud}}{\sqrt{2}}\frac{1}{12}f_Bf_D\sqrt{m_Bm_D} B_1\nonumber \\
k_H&=&-\frac{G_F V_{cb} V_{ud}}{\sqrt{2}}f_{\pi}^2\xi(\omega) B_2 \nonumber \\
k_S&= & 2\frac{G_F V_{cb} V_{ud}}{\sqrt{2}}f_{\pi}^2\tau_{1/2}(\omega) B_2 \nonumber \\
k_T&=&\frac{G_F V_{cb} V_{ud}}{\sqrt{2}}\sqrt{3}\tau_{3/2}(\omega)f_{\pi}^2 B_3.
\label{weakcons}
\end{eqnarray}
In order to treat these weak vertices as a phenomenological parametrization we have introduced ``bag factors'' $B_{1,2,3}$ that are equal to 1 in simple factorization but that we will allow to vary. 
We will also use recent estimates for the IW functions in the light-front formalism \cite{Cheng:2003sm}, taking $\omega \equiv v\cdot v^\prime \approx 1.26$, an average for the range  (1 - 1.53) that occurs in our application. We thus use: $\xi(\omega)\approx 0.7$, $\tau_{1/2}(\omega)\approx 0.25$ and $\tau_{3/2}(\omega)\approx 0.35$. In addition we take $f_B =191$~MeV and $f_D=225$~MeV. 
The relative sign of the constants $\beta_W$ and $k_S$ is important to reproduce the two body decay $B^-\to D_0^0 \pi^-$, as discussed in Ref.~\cite{Jugeau:2005yr}. Here we use the sign implied in Eq.~\ref{weakcons} which reproduces the measured $B^-\to D_0^0 \pi^-$ rate with bag parameters close to one.

Notice that this framework will describe all the non-resonant diagrams in terms of the same coupling constants as the ``quasi-resonant'' diagrams. Our strategy  is thus to fix as many constants as possible from the on-shell two body decay modes $B^- \to D^{\star\star}\pi^-$ and then use these results, supplemented with the HQET description of the strong $D^{\star\star}$ couplings to predict the three-body decay mode. 

\section{$B^- \to D_s^+ K^- \pi^-$ and the $D^{\star\star}$ resonances}

We are now in a position to investigate the contribution of the $D^{\star\star}$ resonances to the $B^- \to D_s^+ K^- \pi^-$ process. Our strategy will be to compute the amplitude with the ingredients given in the previous section, schematically splitting it into the two terms pictured in Figure~\ref{blob}. In the limit in which the quasi-resonant states dominate, it is possible to make reliable predictions that depend only on the theory of the strong $D^{\star\star} D M$ transitions because the weak production vertex and the strong decay vertex factorize, as they do in the narrow width approximation for a resonant channel. The weak transition can then be eliminated in favor of the measured 
$B^- \to D^+ \pi^- \pi^-$. This approach completely fails as the non-resonant contribution becomes dominant in which case the two decay modes are not directly related. The formalism in the previous section will serve to interpolate between these two extremes, allowing us to explore the sensitivity of our result to the weak couplings.

\begin{figure}[htb]
\includegraphics[width=6.5in]{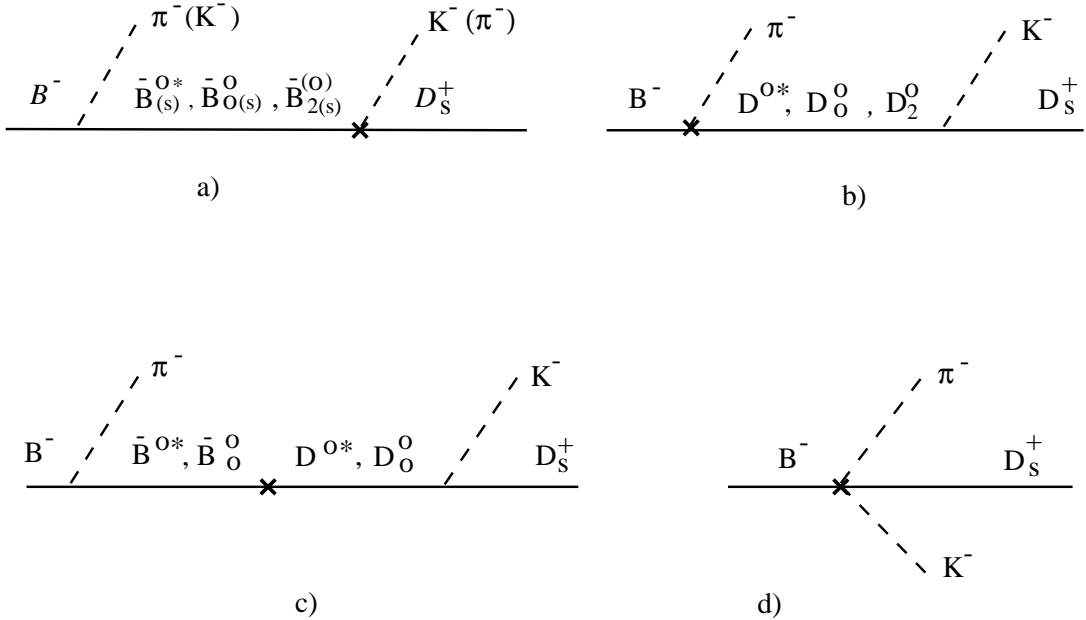}
\centering
\caption{Diagrams contributing to $B^- \to D_s^+ K^- \pi^-$. For diagrams of the form (a), the strange $B$ intermediate states go with a pion emission from the weak vertex (denoted by an x).}
\label{BDKpidiag}
\end{figure}

We begin by computing the $B^- \to D_s^+ K^- \pi^-$ amplitude from the diagrams shown in Fig.\ref{BDKpidiag}. To construct these diagrams we note that weak transitions involving the $T$ multiplet arise only from Eq.~\ref{tenoper}, so the corresponding vertices always involve a $\pi^-$.
The calculation proceeds as follows. We work in the $B$ rest frame in which the heavy meson has four velocity $v^\mu=(1,0,0,0)$. After a weak light meson emission we solve for the velocity of the charmed heavy meson, $v^\prime$ for $D_{0,2}^0,D^{0*}$, using exact kinematics so that the residual momentum in the intermediate heavy meson propagator corresponds exactly to its off-shellness. For a weak vertex with  pion emission we use
\begin{equation}
v^\prime= (\frac{m_B-E_\pi}{M_{D_s^+ K^-}},-\frac{\vec{p}_\pi}{M_{D_s^+ K^-}}),\, \, \omega= v\cdot v^\prime= \frac{m_B^2+M_{D_s^+ K^-}^2}{2m_B M_{D_s^+ K^-}},
\end{equation}
whereas for a weak vertex with a kaon emission we use
\begin{eqnarray}
v^{\prime\prime} =(\frac{m_B-E_K}{M_{D_s^+ \pi^-}},-\frac{\vec{p}_K}{M_{D_s^+ \pi^-}}).
\end{eqnarray}

To determine phenomenological values for our ``bag factors'' we use the experimental results \cite{Abe:2003zm},
\begin{eqnarray}
{\cal B}(B^- \rightarrow D^0_0 \pi^-){\cal B}(D^0_0 \rightarrow D^+ \pi^-)&=& (6.1\pm 0.6 \pm 0.9 \pm 1.6)\times 10^{-4}\nonumber \\
{\cal B}(B^- \rightarrow D^0_2 \pi^-){\cal B}(D^0_2 \rightarrow D^+ \pi^-)&=& (3.4 \pm 0.3 \pm 0.6 \pm 0.4)\times 10^{-4},
\end{eqnarray}
supplemented with the theoretical input for the strong $D^{\star\star}$ decays as in \cite{Jugeau:2005yr,Cheng:2006dm}. For the central values then,
\begin{eqnarray}
{\cal B}(B^- \rightarrow D_0^0 \pi^-)&=& 9.1\times 10^{-4} \nonumber \\
{\cal B}(B^- \rightarrow D_2^0 \pi^-)&=& 8.7 \times 10^{-4}.
\label{prodbr}
\end{eqnarray}
The weak decay rates $B^- \rightarrow D^0_0 \pi^-$ and $B^- \rightarrow D^0_2 \pi^-$  can be calculated in the above framework with results,
\begin{eqnarray}
\Gamma(B^- \rightarrow D^0_0 \pi^-)&=&\frac{E_{D_0^0}E_{\pi}}{8\pi m_Bf_\pi}\left[4\beta_W^\prime \omega_0 \left(1+h\frac{m_B+m_{D_0^0}}{2m_{D_0^0}}\right)-\frac{k_S E_{\pi}(m_{D_0^0}-m_B)}{ m_{D_0^0}}  \right]^2 , \nonumber \\
\Gamma(B^- \rightarrow D^0_2 \pi^-)&=&\frac{k_T^2E_{D_2}E_{\pi}(\omega_2^2-1)^2 (m_B+m_{D_2})^2}{12\pi^2 m_B f_{\pi}^2},
\label{tenprod}
\end{eqnarray}
where $\omega_0=\frac{E_{D_0^0}}{m_{D_0^0}}=1.362$ and $\omega_2=\frac{E_{D_2^0}}{m_{D_2^0}}=1.306$. Setting these predictions, Eq.~\ref{tenprod}, equal to the values in Eq.~\ref{prodbr} 
we find $B_2$ as a function of $B_1$. For our numerics we will use the three pairs $B_1=1$, $B_2=1.13$; $B_1= 1.308$, $B_2=1$ and $B_1=1.15$, $B_2=1.06$. This comparison has also been used to fix the relative sign of $k_S$ with respect to $\beta_W^\prime$. From the second line it also follows that $B_3=1.3$.

We neglect mass splittings between members of a doublet, $H,S,T$, but include mass splittings between the different doublets. In addition, we set the pion mass to zero. We find it convenient to evaluate scalar products involving $v^\prime$ in the $D_s^+ K^-$ center of mass frame.

\begin{figure}[htb]
\includegraphics[width=5in]{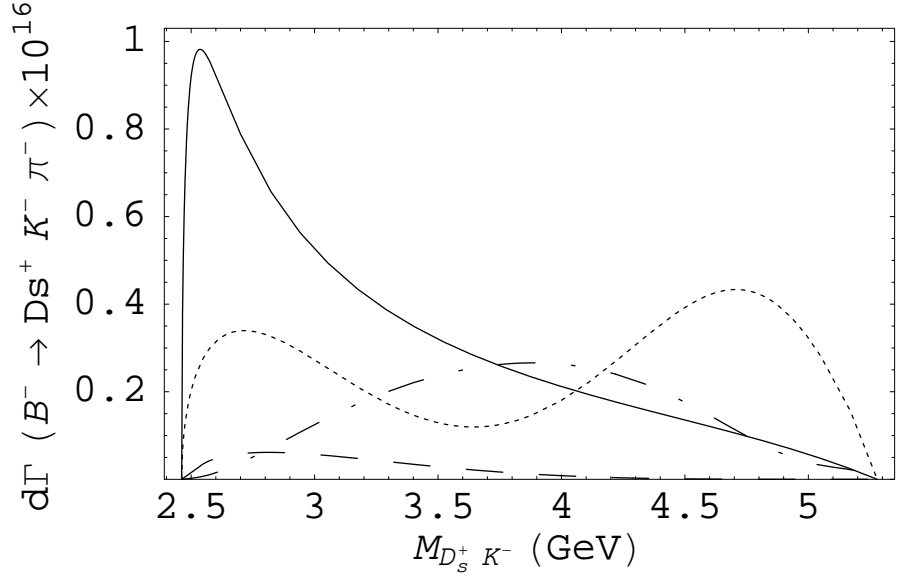}
\centering
\caption{$M_{D_s^+K^-}$ invariant mass distribution for $B_1=1$, $B_2=1.13$ for: a) diagrams involving a $D_0^0$ (solid line); b) diagrams involving a $D_2^0$ (dashed line) and c) all other diagrams: for $h^\prime>0$ (dotted line) and for $h^\prime<0$ (dash-dotted). Interference terms between (a), (b), and (c) are not shown.}
\label{res}
\end{figure}

All this results in the following amplitudes. The three ``quasi-resonant'' diagrams (those that contain a $D_0^0$ meson in Fig.\ref{BDKpidiag}.b,c) give:
\begin{equation}
{\cal M}_{S}=-\frac{ h}{f_{\pi}f_K} \frac{v^\prime\cdot q_K}
{M_{D_s^+K^-}-m_{D_0^0}}  
\left[4\beta_W^\prime \omega \left(1+h\frac{m_B+M_{D_s^+K^-}}{2M_{D_s^+K^-}}\right)-\frac{k_S E_{\pi}(M_{D_s^+K^-}-m_B)}{M_{D_s^+K^-}} \right],
\label{ampres}
\end{equation}
with $E_\pi$ evaluated in the $B$ rest frame. This $D_0^0$ contribution by itself  has an $M_{D_s^+ K^-}$ invariant mass distribution shown as the solid line in Fig.\ref{res}. 

There is one ``quasi-resonant'' diagram with an intermediate tensor $D_2^{\star\star}$ state (Fig.\ref{BDKpidiag}.b). It yields an amplitude,
\begin{eqnarray}
{\cal M}_T=\frac{h^\prime k_T}{f_{\pi}f_K(M_{D_s^+K^-}-m_{D_2^0})} 
\left\{ q_{\pi}\cdot v^\prime \left[ (E_K-\omega q_{K}\cdot v^\prime )^2-\frac{1}{3}(1-\omega^2)(m_K^2-(q_{K}\cdot v^\prime)^2)\right] 
\right. \nonumber \\
-\left. (\omega+1) \left[ (q_{\pi}\cdot q_{K}-q_{\pi}\cdot v^\prime q_{K}\cdot v^\prime )(E_K-\omega q_{K}\cdot v^\prime ) 
-\frac{1}{3} (m_K^2-(q_{K}\cdot v^\prime)^2 )(E_{\pi}-\omega q_{\pi}\cdot v^\prime)\right] \right\}, \nonumber \\
\label{amptensor}
\end{eqnarray}
where $E_{\pi,K}$ are evaluated in the $B$ rest frame. 
By itself the $D_2^0$ contribution  has an $M_{D_s^+ K^-}$ invariant mass distribution shown as the dashed line in Fig.\ref{res}.

Finally there are the ``non-resonant'' diagrams that we divide into two groups. The diagrams from Fig.\ref{BDKpidiag}.a,d give:
\begin{eqnarray}
&{\cal M}_{other}&=-\frac{4\beta_W^\prime \omega}{f_{\pi}f_K} \left( 1-\frac{h q_{\pi}\cdot v^\prime}{M_{D_s^+K^-}-m_B}\right)+\frac{hk_S (v_{D_s^+}-v^{\prime\prime})\cdot q_{\pi}\hspace{1mm}v^{\prime\prime}\cdot q_K}{f_{\pi}f_K(M_{D_s^+\pi^-}-m_{B^0_{0s}})} \label{ampnonres}\\
&+& \frac{4\beta_W^\prime g(q_{\pi}\cdot q_D-q_{\pi}\cdot v^\prime q_{D}\cdot v^\prime)}{f_{\pi}f_K (M_{D_s^+K^-}-m_{B})m_D}\nonumber \\
&-&\frac{gk_H(q_{\pi}\cdot q_K-q_{\pi}\cdot v^{\prime\prime} q_{K}\cdot v^{\prime\prime})}{f_{\pi}f_K (M_{D_s^+\pi^-}-m_{B_s^\star})}(\frac{2q_D \cdot q_{\pi}}{m_Bm_D}+\frac{m_D}{m_B}+1) \nonumber \\
&-& \frac{h^\prime k_T }{f_{\pi}f_K (M_{D_s^+\pi^-}-m_{B_{2s}})}  
\left\{ (v^{\prime\prime} \cdot v_{D_s^+}+1) \left[ (q_{\pi}\cdot q_{K}-q_{\pi}\cdot v^{\prime\prime} q_{K}\cdot v^{\prime\prime} ) \right.\right.\nonumber \\
&\times& \left.(q_{K}\cdot v_{D_s^+}-v^{\prime\prime} \cdot v_{D_s^+} q_{K}\cdot v^{\prime\prime} )-
\frac{1}{3} (m_K^2-(q_{K}\cdot v^{\prime\prime})^2) (q_{\pi}\cdot v_{D_s^+}-v^{\prime\prime} \cdot v_{D_s^+} q_{\pi}\cdot v^{\prime\prime})\right] \nonumber \\
&-& \left. q_{\pi}\cdot v^{\prime\prime} \left[ (q_{K}\cdot v_{D_s^+}-v^{\prime\prime} \cdot v_{D_s^+} q_{K}\cdot v^{\prime\prime} )^2-\frac{1}{3}(1-(v^{\prime\prime} \cdot v_{D_s^+})^2)(m_K^2-(q_{K}\cdot v^{\prime\prime})^2)\right] \hspace{3mm}   \right\}. \nonumber
\end{eqnarray}
Diagrams from Fig.\ref{BDKpidiag}.b,c containing a $D^{0*}$ intermediate state give:
\begin{eqnarray}
{\cal M}_{D^\star}&=&-\frac{4g\beta_W^\prime}{f_{\pi}f_K (M_{D_s^+K^-}-m_{D^\star})} \left[ 1-\frac{k_H}{4\beta_W^\prime} q_{\pi}\cdot v^\prime \right](E_K-\omega q_K \cdot v^\prime)\nonumber\\
&+&\frac{g(q_{\pi}\cdot q_K-q_{\pi}\cdot v^\prime q_{K}\cdot v^\prime)}{f_{\pi}f_K (M_{D_s^+K^-}-m_{D^\star})}\left[-k_H(\omega+1)+\frac{4g\beta^\prime_W}{M_{D_s^+K^-}-m_{B^*}}\right] .
\label{ampnonres1}
\end{eqnarray}
To obtain the partial contribution from the ``non-resonant'' diagrams to the $M_{D_s^+ K^-}$ invariant mass distribution it is necessary to know the sign of  $h^\prime$. Since this sign is not known,  we present results for both signs shown as dotted and dash-dotted lines in Figure~\ref{res}.

\begin{figure}[htb]
\includegraphics[width=2.9in]{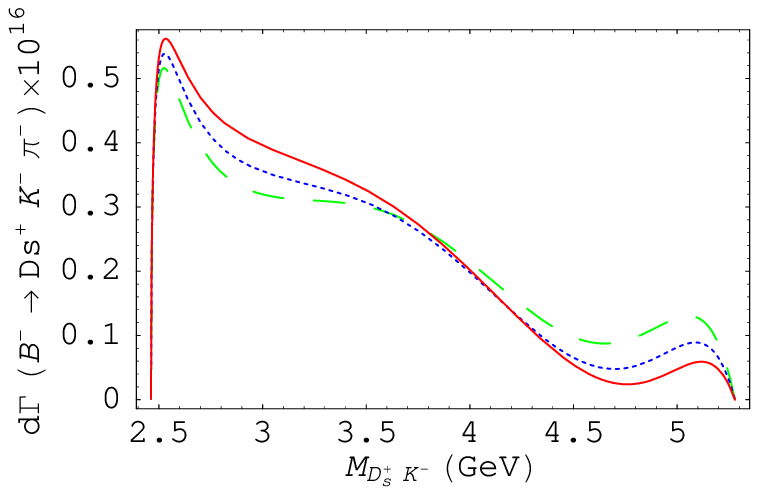} \includegraphics[width=2.9in]{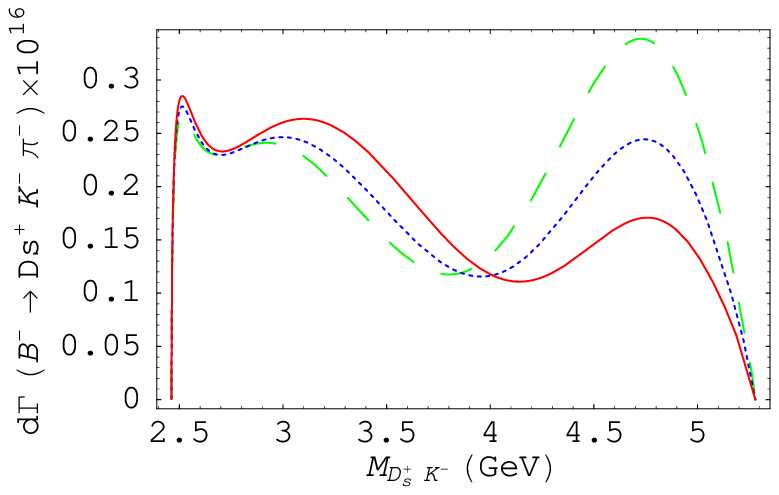}
\centering
\caption{$M_{D_s^+ K^-}$ invariant  mass distributions with $B_1=1$, $B_2=1.13$ (solid), $B_1=1.15$, $B_2=1.06$ (dotted), $B_1=1.308$, $B_2=1$ (dashed) for (a) $h^\prime <0 $ and (b) 
$h^\prime >0$.}
\label{all}
\end{figure}

We now show the full result obtained by adding all contributions. This is not equal to the sum of the three curves in Figure~\ref{res} because that decomposition ignored the interference between different terms.  
With $h^\prime<0$ we obtain a total ${\cal B}(B^- \to D_s^+ K^- \pi^-)=1.63\times 10^{-4}$ and with $h^\prime > 0$ we find ${\cal B}(B^- \to D_s^+ K^- \pi^-)=1.24\times 10^{-4}$. The corresponding $M_{D_s^+ K^-}$ invariant  mass distributions are shown in Fig.\ref{all} for three values of $B_1$ and $B_2$ as described earlier.   

\section{Discussion}

\subsection{Angular distributions}

We now turn our attention to angular distributions and the additional information they provide. In particular, by studying the angular distribution $d\Gamma(B^- \to D_s^+ K^- \pi^-)/d\cos\theta$ for the angle $\theta$ between the momenta of the pion and the kaon in the $D_s^+ K^-$ center of mass frame we can extract the amplitudes with different angular momentum. This frame would correspond to the rest frame of the $D^{\star\star}$ if it were produced as a physical intermediate state, so that this is the angular distribution that would normally be used to determine the spin of the resonance. In $B^- \to D_s^+ K^- \pi^-$ there is no resonance in the physical region, but we expect the different contributions to the rate to exhibit different angular distributions depending on the virtual intermediate state. This is made evident by rewriting the amplitudes in the $D_s^+ K^-$ center of mass frame. We find that the total amplitude in this frame can be written as a linear superposition of   Legendre polynomials in $\cos\theta_{K^-\pi^-}$ (the angle between the $K^-$ and $\pi^-$ momenta in the $D_s^+ K^-$ center of mass frame),
\begin{equation}
{\cal M}(B^- \to D_s^+ K^- \pi^-)={\cal M}_0 P_0(\cos\theta_{K^-\pi^-}) +{\cal M}_1 P_1(\cos\theta_{K^-\pi^-}) +{\cal  M}_2 P_2(\cos\theta_{K^-\pi^-}) + \cdots
\end{equation} 
With sufficient statistics it should be possible to fit the observed 
angular distribution to this form and to extract the different ${\cal M}_i$ components.

In terms of these components we can write the 
differential decay rate as\footnote {The starred energies and momenta are evaluated in the $D_s^+K^-$ center of mass frame.} 
\begin{equation}
\frac{d\Gamma(B^- \to D_s^+ K^- \pi^-)}{dM_{D_s^+K^-}}= \frac{E_{D_s^+} |\vec{p}_K^{~\star}| |\vec{p}_{\pi}|}{4(2\pi)^3m_B}\left[|{\cal M}_0|^2  +\frac{1}{3}|{\cal M}_1|^2 +\frac{1}{5}|{\cal  M}_2|^2 +\cdots\right ],
\end{equation} 
and compare the different contributions, which 
correspond to the $D_s^+K^-$ system having angular momentum 0, 1 or 2 respectively.

\begin{figure}[htb]
\includegraphics[width=2.9in]{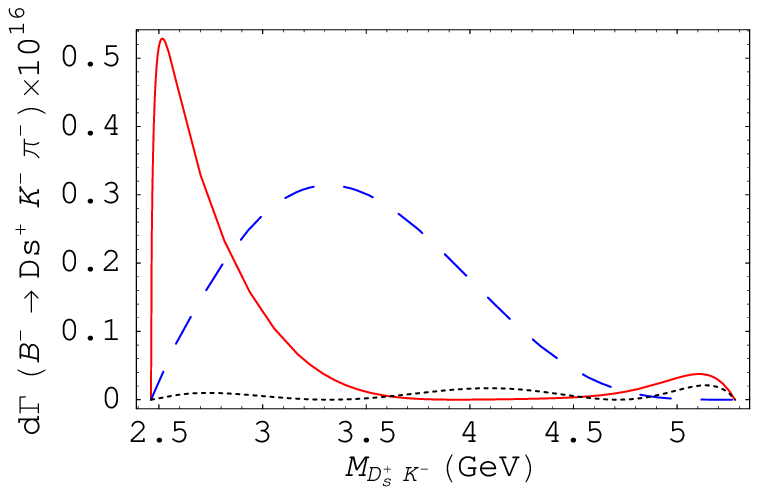} \includegraphics[width=2.9in]{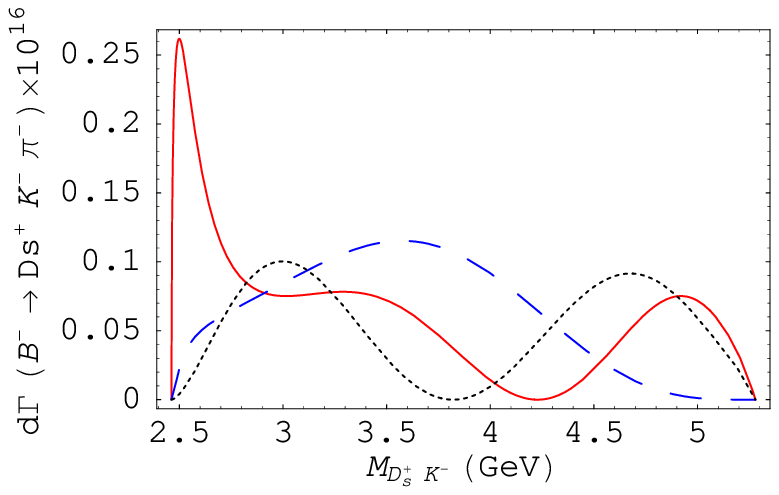}
\centering
\caption{Contributions to the decay rate from different spin amplitudes for $B_1=1$, $B_2=1.13$ with: (a) $h^\prime< 0$, (b) $h^\prime> 0$. In both cases the solid line corresponds to $M_0$, the dotted line to $M_1$ and the dashed line to $M_2$. Higher spin contributions are negligible and are not shown.  }
\label{angdec}
\end{figure}

Within our framework, the partial amplitudes are predicted to be,
\begin{eqnarray}
{\cal M}_0&=& -\frac{ h E_K^\star }{f_{\pi}f_K(M_{D_s^+K^-}-m_{D_0^0})} \left \{ 4\beta_W^\prime\omega\left[1+h\frac{M_{D_s^+K^-}+m_B}{2M_{D_s^+K^-}}  - \frac{E_{\pi}^\star (M_{D_s^+K^-}-m_{D_0^0})}{E_K^\star(M_{D_s^+K^-}-m_B)}\right]
\right. \nonumber \\
&-& \left. k_S E_{\pi}^\star\frac{M_{D_s^+K^-}-m_B}{m_B} \right \} -\frac{4\beta_W^\prime \omega}{f_{\pi}f_K} +a_0 \nonumber \\
{\cal M}_1&=& \frac{g |\vec{q}_{\pi}| |\vec{q}_K|}{f_{\pi}f_K(M_{D_s^+K^-}-m_{D^{*0}})} \left \{ 4\beta_W^\prime \left[\frac{1}{m_B}-\frac{g}{M_{D_s^+K^-}-m_{B^*}}
+ \frac{M_{D_s^+K^-}-m_{D^{*0}}}{(M_{D_s^+K^-}-m_{B^*})m_D} \right]
\right. \nonumber\\
&+& \left. k_H \frac{M_{D_s^+K^-}+m_B}{m_B} \right\}+a_1 \nonumber \\
{\cal M}_2&=& \frac{2}{3} \frac{h^\prime k_T |\vec{q}_{\pi}|^2 |\vec{q}_K|^2}{f_{\pi}f_K}\frac{M_{D_s^+K^-}+m_B}{(M_{D_s^+K^-}-m_{D_2^0})m_B^2}+a_2  \label{pwaves}.
\end{eqnarray}
In Eq.~\ref{pwaves} we have shown explicitly the contributions to the $J=0,1,2$ amplitudes from the diagrams with an intermediate, ``quasi-resonant'', $D_0^0$, $D^{\star 0}$ and $D_2^0$ respectively. Additional contributions arise from the non-resonant diagrams, in particular the diagrams in which $B_{0s}^0$, $B_s^{0*}$ and $B_{2s}^0$ are exchanged with the $K^-$ connected to the $B$ vertex and the $\pi^-$ connected to the $D_s^+$ vertex contribute to all values of angular momentum. We denote their projections into $J=0,1,2$  by $a_0,~a_1,~ a_2$ and evaluate these contributions numerically. We show this decomposition in Fig.\ref{angdec} for $B_1=1$, $B_2=1.13$ as a function of $D_s^+K^-$ invariant mass.

\begin{figure}[htb]
\includegraphics[width=4in]{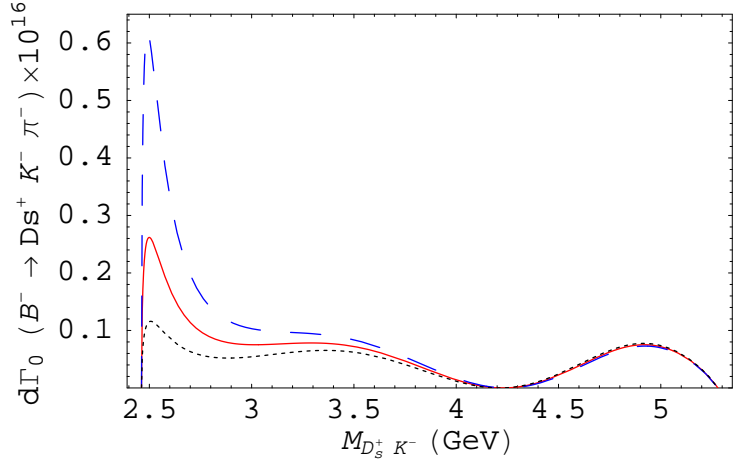} 
\centering
\caption{Scalar component of $\Gamma(B^- \to D_s^+K^-\pi^-)$ for three different values of $m_{D_0^0}$ corresponding to the central value in Eq. ~\ref{measmass} (solid), and to the one standard deviation values (dashed, dotted), with $h^\prime >0$.}
\label{3masses}
\end{figure}

In Figure~\ref{angdec} we see that the contribution from $M_0$ is peaked at low $M_{D_s^+K^-}$ and that the height of the peak depends on the sign of $h^\prime$ through the interference between quasi-resonant and non-resonant diagrams. It is larger for $h^\prime <0$ where the non-resonant background is smaller in the low $M_{D_s^+K^-}$ region as seen in Figure~\ref{res}. This peak reflects the presence of the $D_0^0$ resonance just outside the physical region, and its height is sensitive to the mass of the $D_0^0$, as illustrated in Figure~\ref{3masses}. In that figure we show the result with the central value of Eq.~\ref{measmass} as a solid line, and the one standard deviation values (adding all errors in quadrature) of $m_{D_0^0}$ as dashed and dotted lines.    

Figure~\ref{angdec} also shows a large difference between the size of the $M_1$ contribution for different signs of $h^\prime$ and this may be exploited to determine this sign. Similarly, the contribution from $M_2$ is significantly larger when $h^\prime <0$ providing another handle on determining this sign.
To further quantify the different contributions we show the respective partial branching ratios in Table \ref{table}.

\begin{table}
\caption{Partial branching ratios for spin amplitudes.}
\label{table}
\begin{center}
\begin{tabular}{|l|l|l|}
\hline
&$\hspace{2mm}h^\prime>0\hspace{2mm}$&$\hspace{2mm}h^\prime<0\hspace{2mm}$\\
\hline
$\hspace{2mm}{\cal B}_0(B^-\to D_s^+ K^-\pi^-)\hspace{2mm}$&$\hspace{2mm}4.4\times 10^{-5}\hspace{2mm}$& $\hspace{2mm}5.3 \times 10^{-5}\hspace{2mm}$ \\
$\hspace{2mm}{\cal B}_1(B^-\to D_s^+ K^-\pi^-)\hspace{2mm}$ &$\hspace{2mm}3.6 \times 10^{-5}\hspace{2mm}$& $\hspace{2mm}5.6 \times 10^{-6}\hspace{2mm}$\\
$\hspace{2mm}{\cal B}_2(B^-\to D_s^+ K^-\pi^-)\hspace{2mm}$&$\hspace{2mm}4.3 \times 10^{-5}\hspace{2mm}$& $\hspace{2mm}1.0 \times 10^{-4}\hspace{2mm}$\\
$\hspace{2mm}{\cal B}_{J>2}(B^-\to D_s^+ K^-\pi^-)\hspace{2mm}$&$\hspace{2mm}4.2 \times 10^{-8}\hspace{2mm}$ &$\hspace{2mm}5.4 \times 10^{-7}\hspace{2mm}$\\
\hline
\end{tabular}
\end{center}
\end{table}

\subsection{Dependence on the parametrization of the weak vertex}

\begin{figure}[htb]
\includegraphics[width=2.9in]{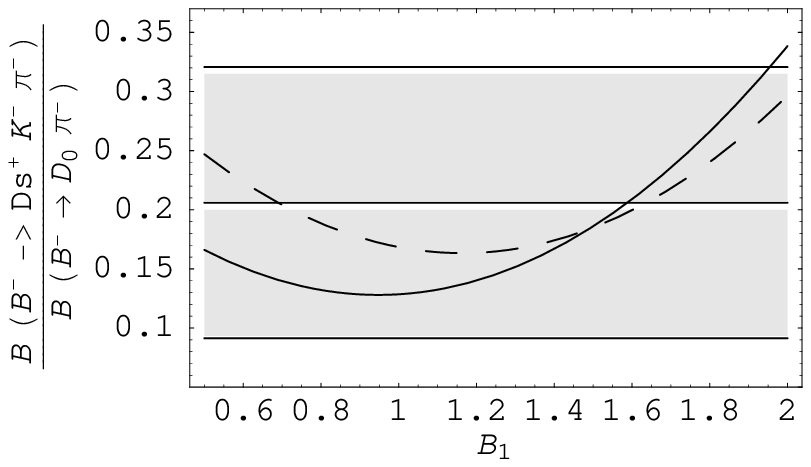} \includegraphics[width=2.9in]{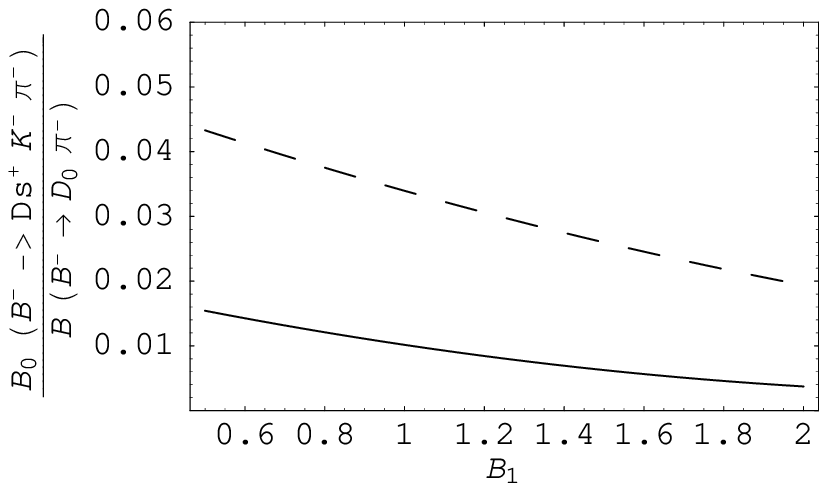}
\centering
\caption{Normalized $\Gamma(B^-\to D_s^+K^-\pi^-)$ as a function of $B_1$ factor for: a) full kinematic range; and b) scalar contribution in the $M_{D_s^+K^-}\leq m_{D^0_0}+2\Gamma_{D^0_0} \sim 2.8$~GeV range.  The horizontal lines in (a) show the 1-$\sigma$ range from the BaBar measurement \cite{batalk}. In both cases the solid line corresponds to $h^\prime >0$ and the dashed line to $h^\prime <0$.}
\label{B1}
\end{figure}

In Figure~\ref{all} we have already presented results for three different pairs of values for $B_1$ and $B_2$. These correspond to different parametrizations for the weak vertex that reproduce the central value of the two body decay rates. We see from that figure that the variations are not large in the low $M_{D_s^+ K^-}$ region. We now explore in more detail the dependence of the total rate on these parameters.  To this end we first normalize the total decay rate $\Gamma(B^-\to D_s^+K^-\pi^-)$ to the rate $\Gamma(B^- \rightarrow D^0_0 \pi^-)$ calculated in Eq.~\ref{tenprod}. We then plot this ratio as a function of the ``bag factor'' $B_1$ while adjusting $B_2$ in such a way that $B^- \rightarrow D^0_0 \pi^-$ remains fixed to its   experimental  (central)  value in Figure~\ref{B1}a.  

We see that the ratio changes by about a factor of two when we span the value of $B_1$ from $1/2$ to $2$ (recall $B_1=1$  in naive factorization). This variation indicates that our prediction for the full rate ${\cal B}(B^- \to D_s^+ K^- \pi^-)$ is not robust over the full kinematic range and that the process is not dominated by the $D_0^{\star\star}$ quasi-resonance. In the figure we also show the 1-$\sigma$ band from the BaBar measurement \cite{batalk} and we see that our prediction for the total rate is in good agreement.

Following our general discussion, we might expect to do better if we limit the range for $M_{D_s^+K^-}$ to values closer to the physical $m_{D_0^0}$ mass, since this would enhance the relative contribution of the ``quasi-resonance''. For illustration we repeat the above exercise including only the partial branching ratio ${\cal B}_0(B^- \to D_s^+ K^- \pi^-)$ from the region $M_{D_s^+K^-}\leq m_{D^0_0}+2\Gamma_{D^0_0} \sim 2.8$~GeV.  We also limit the comparison to the scalar contribution to the rate, as this is the one that could be dominated by the $D_0^0$ quasi-resonance. We show these results in Fig.\ref{B1}b. We notice a slight improvement in the form of reduced dependence of our prediction on the parametrization of the weak vertex. However, there is still a large dependence on the parametrization of the weak vertex as the ratio varies by a factor of about two in the range $0.5<B_1<2$. This is not surprising as Figure~\ref{angdec} already showed that there is a large non-resonant contribution present.

\section{$B^- \to D_{0,2}^0 \pi^- \to D^{0*}\gamma \pi^-$}

We end this paper with a brief  discussion of  the radiative decays $D_{0,2}^0 \rightarrow D^*_0\gamma$. The framework used in this paper should be more reliable for the decay chains $B^- \to D_{0,2}^0 \pi^- \to D^*_0\gamma \pi^-$ because in this case the resonance is in the physical region and should dominate the amplitude. In this case  the approximation in which the weak production of $D_{0,2}^0$ and the subsequent radiative decay factorize should be more reliable.

The production rates $B^- \to D_{0,2}^0 \pi^-$  were already obtained in Eq.\ref{prodbr}. The  radiative decays of the $D_{0,2}^0$ can be readily extracted from the vertices in Ref.~ \cite{Antipin:2006dn}  (see Eqs.~13,14,15,A4 in that reference). We find 
\begin{eqnarray}
\Gamma(D_0^0 \rightarrow D^{0*} \gamma)=(e\mu_D^S)^2 \cdot \frac{E_{\gamma}^3 }{m_{D^0_0}} \cdot \frac{E_{D^{*0}}}{4\pi} \approx 0.09MeV, \nonumber\\
\Gamma(D_2^0 \rightarrow D^{0*} \gamma)=(e\mu_D^T)^2 \cdot \frac{E_{\gamma}^3 }{m_{D^0_2}} \cdot \frac{E_{D^{*0}}}{\pi} \approx 0.20MeV. 
\label{2bodyrad}
\end{eqnarray}
Numerically we used the results $\mu_D^S=\frac{2e_c\tau^{1/2}(1)}{m_c}+\frac{e_u}{\Lambda_{1/2}^\prime}\approx 0.75 GeV^{-1}$ and $\mu_D^T=\frac{e_c\tau^{3/2}(1)}{m_c}+\frac{e_u}{\Lambda_{3/2}^\prime}\approx 0.33 GeV^{-1}$ with input 
parameters discussed in Ref.~ \cite{Antipin:2006dn}.

Using $\Gamma_{D_0^0}=276$~MeV and $\Gamma_{D_2^0}=45$~MeV we find ${\cal B}(D_0^0 \rightarrow D^{0*} \gamma)=3.3 \times 10^{-4}$ and ${\cal B}(D_2^0 \rightarrow D^{0*} \gamma)=4.5 \times 10^{-3}$. With sufficient statistics to observe these modes it will then be possible to extract the coupling constants $\mu_D^S$ and $\mu_D^T$. We present the result for $D_2^0$ for completeness, as it has already appeared in the literature \cite{Korner:1992pz}. 

\section{Conclusions}

We have analyzed the mode $B^- \to D_s^+ K^- \pi^-$ using HQET to parametrize ``quasi-resonant'' and non-resonant contributions. With the aid of angular analysis it should be possible to extract the contributions of virtual intermediate states with spin $0,~1,~{\rm or}~2$ leading to the $D_s^+K^-$ final state. 

The spin zero partial rate receives a large but not dominant contribution from the $D_0^0$ intermediate state. This means that it is not possible to test the $D_0^0 D_s^+ K^-$ vertex in a model independent way. However, we have seen that the HQET description gives a picture for this decay that can be tested qualitatively at least. The shape of the $M_{D_s^+ K^-}$ distribution for this partial rate depends both on the precise value of the $D_0^0$ mass, as well as on the relative size of the non-resonant contributions.

The spin one contribution can be almost as large as the spin zero contribution if $h^\prime > 0$. This is quite surprising as this is  dominated by an intermediate $D^{\star 0}$ with a mass significantly below threshold. This reinforces the conclusion that the decay is not dominated by the $D^{\star\star}$ resonances although they play an important role.  A determination of this contribution should provide strong evidence for the sign of $h^\prime$.

The contribution of spin two is also very large and strongly dependent on the sign of $h^\prime$ indicating that it is not saturated by the ``quasi-resonant'' $D_2^0$ state.  Contributions of spins higher than two are negligible.

Given the large $m_b-m_c$ mass difference we do not expect the momentum expansion to describe this weak decay quantitatively. In particular the pion and kaon in the non-resonant diagrams are not soft for most of the kinematically allowed range. However, we have provided a mixed framework that uses HQET to describe strong transitions and a naive factorization to describe weak transitions in terms of a few phenomenological parameters. This framework also provides a model for the non-resonant terms and produces  a qualitative description for this decay mode that can be used to compare with experiment and to extract information on the signs and $SU(3)$ properties of the strong couplings. 

\begin{acknowledgments}

This work  was supported in part by DOE under contract 
number DE-FG02-01ER41155. We thank Vitaly Eyges, Soeren Prell and Hai-Yang Cheng for useful discussions.

\end{acknowledgments}

\end{document}